\documentclass{PoS}
\newcommand{\be}{\begin{eqnarray}}
\newcommand{\ee}[1]{\label{#1}\end{eqnarray}}
\newcommand{\bfn}{\begin{figure}[htb]\begin{center}}
\newcommand{\efn}[1]{\label{#1}\end{center}\end{figure}}
\newcommand{\btn}{\begin{table}[htb]\begin{center}}
\newcommand{\etn}[1]{\label{#1}\end{center}\end{table}}
\newcommand{\bm}[1]{\mbox{\boldmath $#1$}}
\newcommand{\bms}[1]{\mbox{\scriptsize\boldmath $#1$}}
\newcommand{\rmt}[1]{\tiny\rm #1}
\newcommand{\lw}[1]{\smash{\lower 1.5ex\hbox{#1}}}

\title{Three -Pion Correlations}

\ShortTitle{Three -Pion Correlations}

\author{\speaker{Minoru Biyajima}\\ 
        Department of Physics, Shinshu University, Matsumoto 390-8621, Japan\\
        E-mail: \email{biyajima@azusa.shinshu-u.ac.jp}}

\author{Takuya Mizoguchi\\
        Toba National College of Maritime Technology, Toba 517-8501, Japan\\            E-mail: \email{mizoguti@toba-cmt.ac.jp}}

\author{Naomichi Suzuki\\
        Department of Comprehensive Management, Matsumoto University, Matsumoto 390-1295, Japan\\        
        E-mail: \email{suzuki@matsu.ac.jp}}

\abstract{First of all, we mention the situation of empirical analyses on 3rd order BEC (Bose-Einstein Correlation) at RHIC. Second, we introduce several theoretical formulae / approaches. Third we present our analyses of data in Au+Au at 130 GeV by STAR and preliminary data in Au+Au at 200 GeV by PHENIX Collaborations. Our results also contain analyses by means of core-halo model. Finally, we estimate that the volume of interaction in Au + Au collisions at 130 GeV is 500 fm$^3$, which is compared with $V = R_{\rm long} R_{\rm out} R_{\rm side} \sim$ 300 fm$^3$ in Pb + Pb collision at  2.76 TeV by ALICE Collaboration. Moreover, usefullness of empirical analyses on $(2\pi^+)\pi^-$ and $(2\pi^-)\pi^+$ combinations at RHIC and LHC energies is remarked.}

\FullConference{The Seventh Workshop on Particle Correlations and Femtoscopy\\
		 September 20 - 24 2011\\
		 University of Tokyo, Japan}

\begin{document}

\vspace{-2mm}
\section{Situation of Empirical analyses on 3rd order BEC in Au+Au collision at RHIC}
\vspace{-2mm}
As shown in Table \ref{tb01}, STAR and PHENIX Coll. have reported their analyses at 130 GeV and 200 GeV, respectively. In their analyses, STAR Coll.~\cite{Adams:2003vd} used all combinations of three momentum-transfers $\sqrt{q_{ij}^2} \le Q_3$ (Inside of a globe in Fig. \ref{fg01}),
\be
Q_{inv,3}^2 = q_{12}^2 + q_{23}^2 + q_{31}^2.
\ee{eq01}

On the other hand, PHENIX Coll.~\cite{Csanad:2005nr} used data on diagonal line of a cube in Fig. \ref{fg01},
\be
q_3 = \langle q_{12}\rangle = \langle q_{23}\rangle = \langle q_{31}\rangle\quad {\rm i.\,e.}\quad Q_{inv,3}^2 = 3q_3^2.
\ee{eq03}
Of course this relation holds, as the number of data increases. $\langle \cdots\rangle$ is an average value.
\btn
\vspace{-3mm}
\caption{Situation of analyses on the 3rd order BEC. Notice two empty columns.}
\begin{tabular}{c|c|c}
\hline
$\sqrt{s_{NN}}$  & STAR & PHENIX\\
\hline
130 GeV  & raw and corrected data by $Q_{inv}$\\ 
\hline
200 GeV  &  & preliminary raw and corrected data by $q_3$\\
\hline
\end{tabular}
\vspace{-6mm}
\etn{tb01}
\bfn
\vspace{-3mm}
\resizebox{0.20\textwidth}{!}{\includegraphics{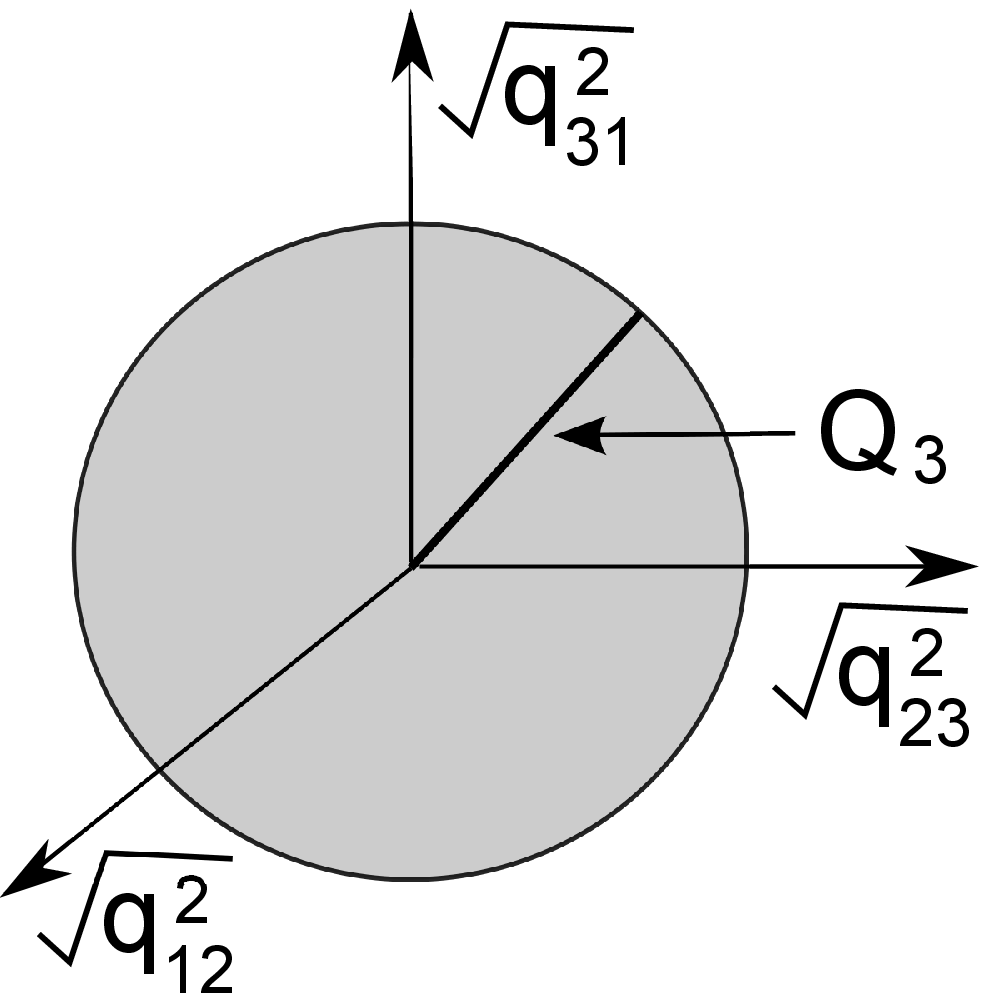}}\hspace{30mm}
\resizebox{0.20\textwidth}{!}{\includegraphics{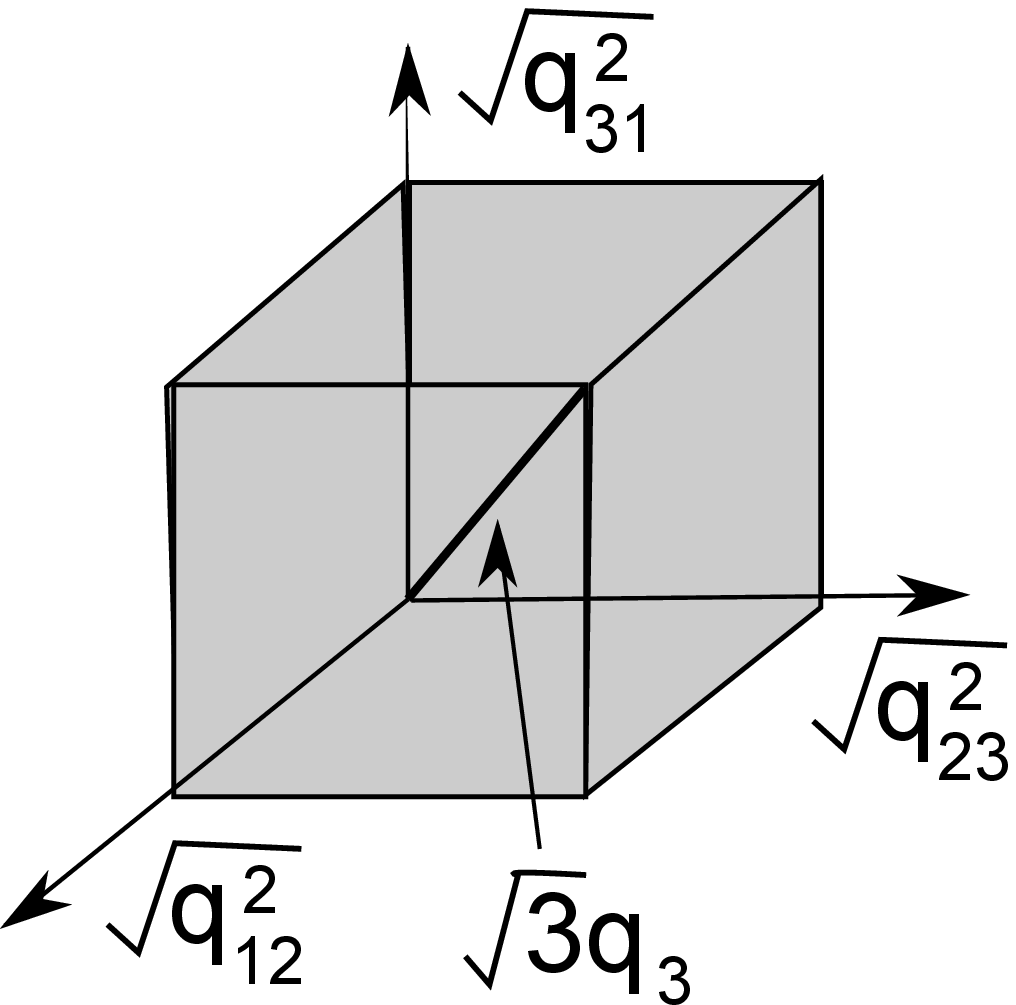}}
\vspace{-3mm}
\caption{Data ensembles of STAR (left) and PHENIX Coll (right).}
\vspace{-6mm}
\efn{fg01}

Here we compare two kinds of data. In Figs.~\ref{fg02}, data by PHENIX Coll.~are rearranged by $\sqrt 3q_3$. Coincidence among data by STAR and PHENIX Coll is fairly good. Error bars in raw data by PHENIX Coll are smaller than those of STAR Coll.
\bfn
\vspace{-3mm}
\resizebox{0.39\textwidth}{!}{\includegraphics{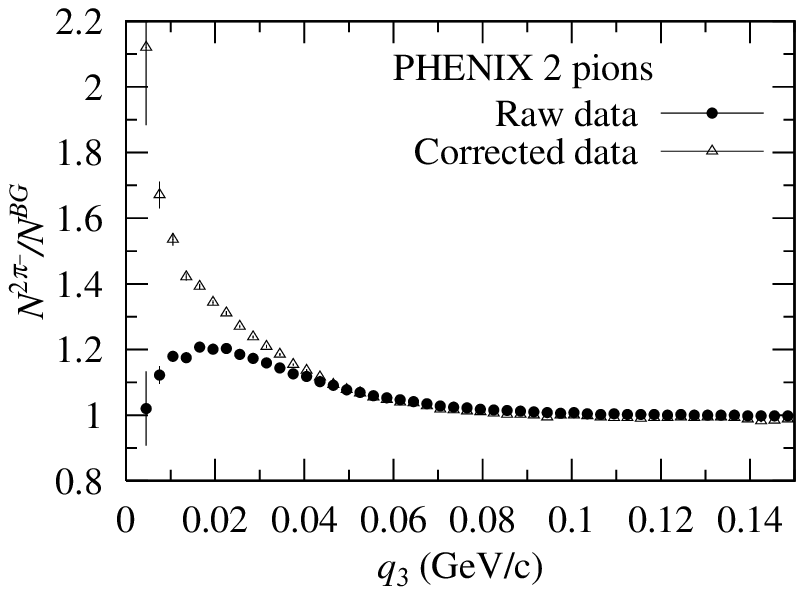}}
\resizebox{0.39\textwidth}{!}{\includegraphics{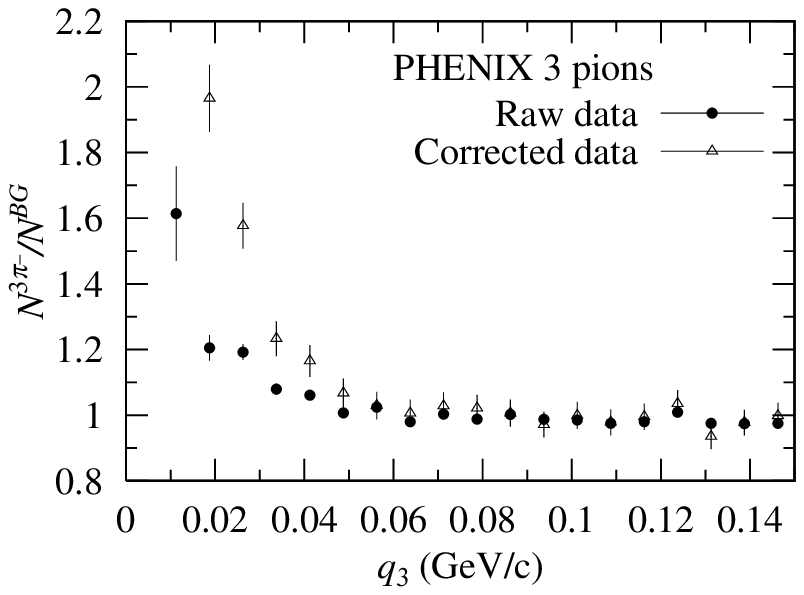}}\\
\resizebox{0.38\textwidth}{!}{\includegraphics{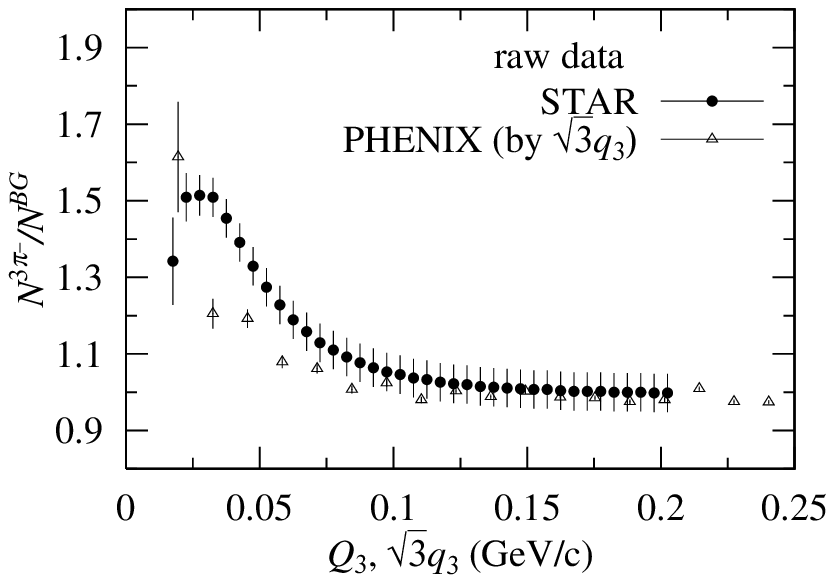}}\quad
\resizebox{0.38\textwidth}{!}{\includegraphics{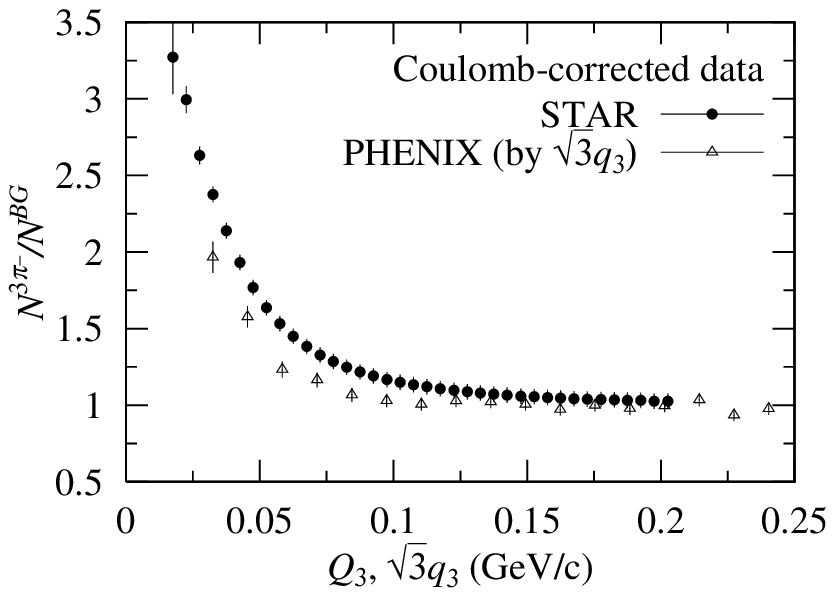}}
\vspace{-3mm}
\caption{Comparisons of data by STAR and PHENIX Coll.}
\vspace{-6mm}
\efn{fg02}

\vspace{-2mm}
\section{Several theoretical formulae}
\vspace{-2mm}
In many analyses on BEC, the following formulae based on plane wave function are used.
\be
  && N^{(2\pi)}/N^{BG} = c\left[1 + \lambda e^{-(RQ)^2}\right],
\label{eq04}\\
  && N^{(3\pi)}/N^{BG} = c\left[1 + \lambda \sum_{i>j} e^{-(RQ_{ij})^2} + 2\lambda^{1.5} e^{-0.5(RQ_3)^2}\right].
\ee{eq05}

In laser optical (LO/GL) approach, the following formulae with a degree of chaoticity $p$ have been proposed \cite{Biyajima:1990ku} and utilized,
\be
N^{(2\pi)}/N^{BG} =\!\!\!&\!\!\!&  1 + 2p(1-p)E_{2B} + p^2E_{2B}^2,
\label{eq06}\\
N^{(3\pi)}/N^{BG} = \!\!\!&\!\!\!&1 + 6p(1-p)E_{3B}
 + 3p^2(3-2p)E_{3B}^2+ 2p^3E_{3B}^3,
\ee{eq07}
where $E_{2B}^2 = \exp(-R^2Q^2)$ (Gaussian form) and/or $E_{2B}^2 = \exp(-R\sqrt{Q^2})$ (exponential form), and  $E_{3B}^3 = \exp(-R^2Q_3^2)$ and so on.

Third we explain the formulae by Coulomb wave function including the degree of coherence $\lambda$ and the interaction range $R$~\cite{Mizoguchi:2000km,Biyajima:2003ey,Biyajima:2005fk,Biyajima:2005ts}. The two-body Coulomb wave function is well known as,
\be
   \psi_{\bms k_i \bms k_j}^C(\bm x_i\bm x_j) = \Gamma(1 + i\eta_{ij})
   e^{\pi \eta_{ij}/2} e^{ i\bms k_{ij} \cdot \bm r_{ij} } 
   F[- i \eta_{ij},\,1;\,i ( k_{ij} r_{ij} - \bms k_{ij} \cdot \bm r_{ij} )],
\ee{eq08}
where, $\bm r_{ij} = \bm x_i - \bm x_j$, $\bms k_{ij} = ( {\bms k_i} - {\bms k_j})/2$, $r_{ij} = |\bm r_{ij}|$, $k_{ij} = |\bms k_{ij}|$ and $\eta_{ij} = e_ie_j\mu_{ij}/k_{ij}$. $\mu_{ij}$ : reduced mass of $m_i$ and $m_j$, $F[a,\ b;\ x]$ : confluent hypergeometric function, $\Gamma(x)$ : Gamma function.

Using of Eq. (\ref{eq08}), the 2nd order BEC with $\lambda$ and Gaussian form for $\rho (x_i)$ is calculated as,
\be
   N^{(2\pi^-)} &\!\!\!=&\!\!\! \frac{1}{2} \prod_{i=1}^2 \int \rho({\bm x_i}) d^3 {\bm x_i}  
     | \psi_{\bm k_1\bm k_2}^C(\bm x_1,\ \bm x_2) 
      + \psi_{\bm k_1\bm k_2}^C(\bm x_2,\ \bm x_1) |^2 \nonumber\\
  &\!\!\!=&\!\!\! \prod_{i=1}^2 \int \rho({\bm x_i}) d^3 {\bm x_i}  
     \left[
     \frac{1}{2} \left( \large| \psi_{\bm k_1\bm k_2}^C(\bm x_1,\ \bm x_2) \large|^2 + \large|\psi_{\bm k_1\bm k_2}^C(\bm x_2,\ \bm x_1) \large|^2  \right)\right.\nonumber\\
    &\!\!\!&\!\!\! \qquad\qquad\qquad\qquad \left.+ \lambda {\rm Re}\left(  \psi_{\bm k_1\bm k_2}^C(\bm x_1,\ \bm x_2) 
      \psi_{\bm k_1\bm k_2}^{C*}(\bm x_2,\ \bm x_1)   \right).
     \right], 
\ee{eq11}

The 3rd order BEC is computed based on the 3-body Coulomb wave function in \cite{Biyajima:2005fk},
\be
  \Psi_f =  \psi_{{\bm k}_1 {\bm k}_2}^{C^{\prime}}({\bm x}_1,{\bm x}_2)
           \psi_{{\bm k}_2 {\bm k}_3}^{C^{\prime}}({\bm x}_2,{\bm x}_3)
           \psi_{{\bm k}_3 {\bm k}_1}^{C^{\prime}}({\bm x}_3,{\bm x}_1),
\ee{eq12}
Hereafter, we use the following expression; $\psi_{{\bm k}_i {\bm k}_j}^{C^{\prime}}({\bm x}_i,{\bm x}_j) = e^{i(2/3){\bm k}_{ij}{\bm r}_{ij}} \phi_{{\bm k}_{ij}}({\bm r}_{ij})$. Notice that the numerical factor ``2/3'' in the exponential function is important~\cite{Biyajima:2005fk,Biyajima:2005ts}.
\be
  N^{(3\pi^-)} &=& \frac{1}{6} \prod_{i=1}^3 \int \rho({\bm x_i}) d^3 {\bm x_i}
  \bigl| \sum_{j=1}^6 A(j) \bigr|^2,  
\ee{eq13}
where 
\vspace{-9mm}
\be
  A(1)&=& A_1= 
  \psi_{\bm k_1\bm k_2}^{C^{\prime}}(\bm x_1,\ \bm x_2)
  \psi_{\bm k_2\bm k_3}^{C^{\prime}}(\bm x_2,\ \bm x_3)
  \psi_{\bm k_3\bm k_1}^{C^{\prime}}(\bm x_3,\ \bm x_1),  \nonumber\\
 A(2)&=& A_{23} =
     \psi_{\bm k_1\bm k_2}^{C^{\prime}}(\bm x_1,\ \bm x_3) 
     \psi_{\bm k_2\bm k_3}^{C^{\prime}}(\bm x_3,\ \bm x_2) 
     \psi_{\bm k_3\bm k_1}^{C^{\prime}}(\bm x_2,\ \bm x_1).
\ee{eq14}
\bfn
\vspace{-3mm}
\resizebox{0.45\textwidth}{!}{\includegraphics{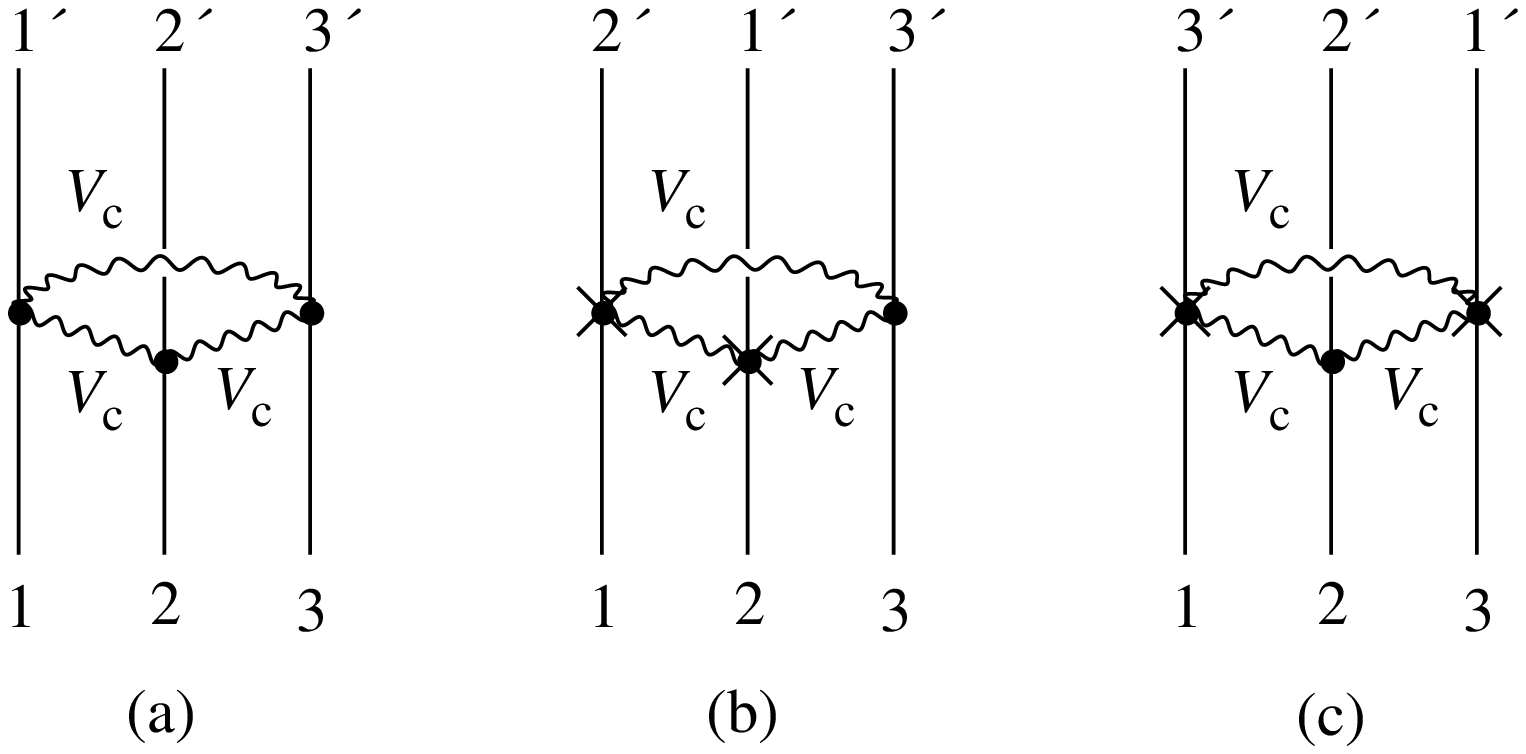}}\qquad
\resizebox{0.45\textwidth}{!}{\includegraphics{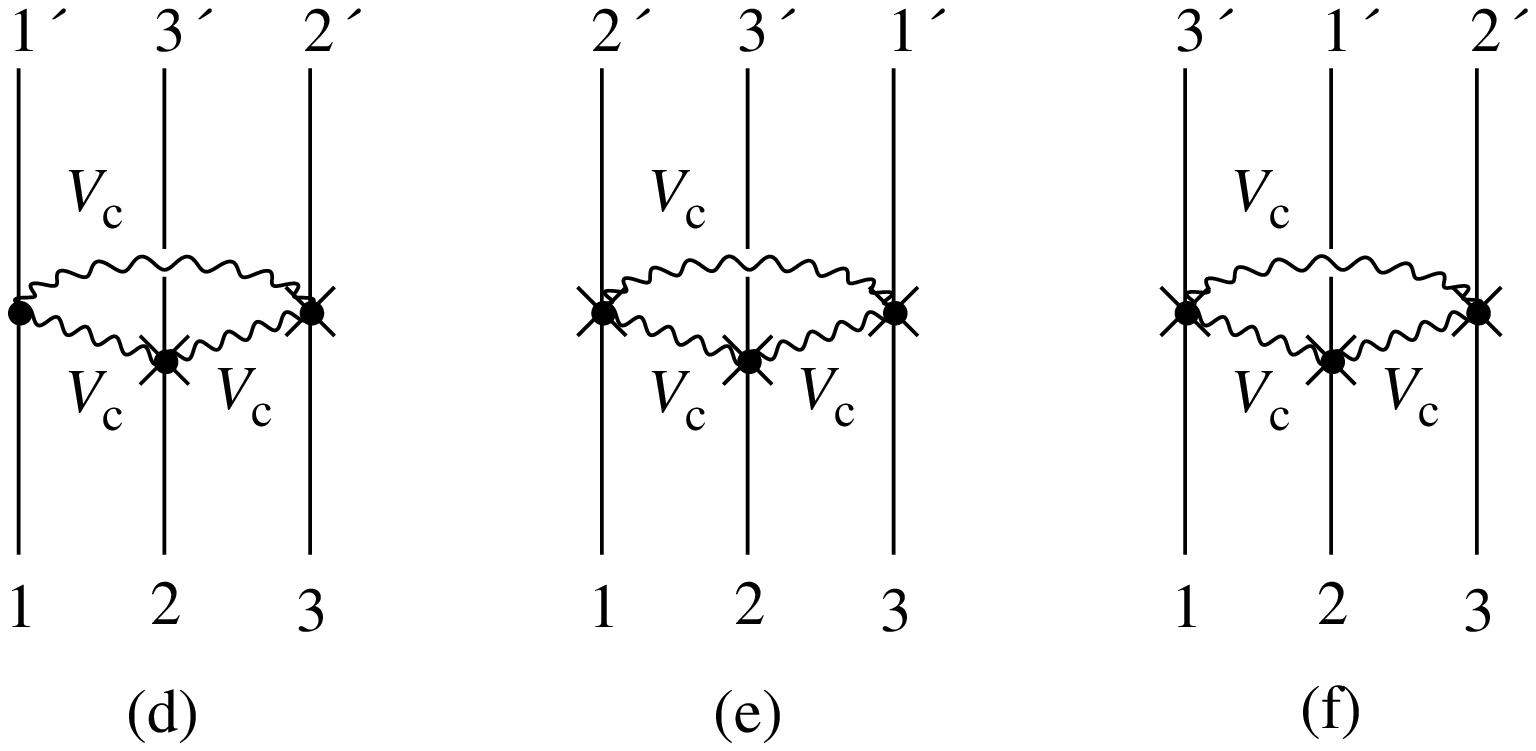}}
\vspace{-3mm}
\caption{Diagrams of the 3rd order BEC for $A(1)\sim A(6)$.}
\vspace{-6mm}
\efn{fg04a}
$A_{ijk}$ is reflecting the permutations of particles, $i,\, j,\, k$ in Fig. \ref{fg04a}: Therein $A(3) = A_{12}$, $A(4) = A_{123}$, $A(5) = A_{132}$ and $A(6) = A_{13}$. In the plane wave approx., we have the correct expression, where $A(3)\sim A(6)$ are skipped \cite{Biyajima:2003ey,Biyajima:2005ts},
\be
 A(1)&=& A_1 \stackrel{\rmt{PW}}{\longrightarrow}
  e^{ i(2/3)( \bm k_{12} \cdot \bm r_{12} 
             +\bm k_{23} \cdot \bm r_{23} 
             +\bm k_{31} \cdot \bm r_{31} )}
  = e^{ i (\bm k_1 \cdot \bm x_1
          + \bm k_2 \cdot \bm x_2 + \bm k_3 \cdot \bm x_3)}, \nonumber \\
 A(2)&=& A_{23} \stackrel{\rmt{PW}}{\longrightarrow} 
    e^{ i(2/3)( \bm k_{12} \cdot \bm r_{13}  
               +\bm k_{23} \cdot \bm r_{32} 
               +\bm k_{31} \cdot \bm r_{21} )}
    = e^{i (\bm k_1 \cdot \bm x_1 
          + \bm k_2 \cdot \bm x_3 + \bm k_3 \cdot \bm x_2)}.
\ee{eq15}
Combining $A(1)\sim A(6)$ in Fig. \ref{fg04a}, we obtain $F_1$ as: (The other formulae $F_{12}\sim F_{123}$ and $F_{132}$ are given in \cite{Biyajima:2005fk,Biyajima:2005ts}.) 
\be
  &\!\!\!&\!\!\! F_1 = \frac{1}{6} [ A_{1}A^*_{1} + A_{12}A^*_{12} + A_{23}A^*_{23}
          + A_{13}A^*_{13} + A_{123}A^*_{123} + A_{132}A^*_{132} ].
\label{eq16}\\
  &\!\!\!&\!\!\! \frac{N^{3\pi^-}}{N^{BG}} = C \prod_{i=1}^3  \int \rho({\bm x_i}) d^3 {\bm x_i}
  \left[ F_1 + 3\lambda F_{12} + 2\lambda^{\frac 32} {\rm Re}\,(F_{123})\right]\nonumber\\
  &\!\!\!&\!\!\! = \frac{C}{(2\sqrt{3}\pi R^2)^3}\int d^3\bm{\zeta}_1d^3\bm{\zeta}_2\exp\left[-\frac 1{2R^2}\left(\frac 12 \bm{\zeta}_1^2 + \frac 23 \bm{\zeta}_2^2\right)\right] 
\left[F_1 + 3\lambda F_{12} + 2\lambda^{\frac 32} {\rm Re}\,(F_{123})\right].
\ee{eq18}
where ${\bm \zeta}_1 = {\bm x}_2 - {\bm x}_1$, ${\bm \zeta}_2 = {\bm x}_3-({m_1{\bm x}_1+m_2{\bm x}_2})/{M_2}$, ${\bm \zeta}_3 = ({m_1{\bm x}_1+m_2{\bm x}_2+m_3{\bm x}_3})/{M}$, $M_2 = m_1 + m_2$ and $M=m_1+m_2+m_3$.

Finally, it is known that the core-halo model is a useful model. However, explicit expressions are skipped here. See our studies in \cite{Mizoguchi:2000km,Biyajima:2003ey,Biyajima:2005fk,Biyajima:2005ts}. See also \cite{Csorgo:1998tn}.
\vspace{-2mm}
\section{Analyses of data by STAR Coll and PHENIX Coll}
\vspace{-2mm}
{\bf 3-1)} For Coulomb corrected data, we employ the conventional formulae, Eqs.~(\ref{eq04}) and (\ref{eq05}). Ours are given in Table \ref{tb02} and Fig. \ref{fg05}. It is interesting that $R_{2\pi} \sim R_{3\pi} \sim$ 8.5 fm for data by STAR Coll.
\btn
\vspace{-3mm}
\caption{Analyses of corrected data by STAR and PPHENIX Coll (Eqs. (\protect \ref{eq04}\protect) and (\protect \ref{eq05}\protect).)}\smallskip
\begin{tabular}{c|ccc||ccc}
  \hline
  && STAR &&& PHENIX\\
  &   $R$ [fm]     &   $\lambda$   &  $\chi^2/{\rm n.d.f.}$  & $R$ [fm] & $\lambda$ & $\chi^2/N_{dof}$\\
  \hline
  $2\pi^-$ & 8.75$\pm$0.31 & 0.58$\pm$0.02 & 23.0/25 & 4.77$\pm$0.04 & 0.39$\pm$0.01 &  178/40 \\
  \hline 
  $3\pi^-$ & 8.26$\pm$0.39 & 0.50$\pm$0.02 & 1.88/35 & 6.92$\pm$0.82 & 0.34$\pm$0.10 &  6.5/14 \\   
  \hline 
\end{tabular}
\vspace{-6mm}
\etn{tb02}
\bfn
\vspace{-3mm}
\resizebox{0.43\textwidth}{!}{\includegraphics{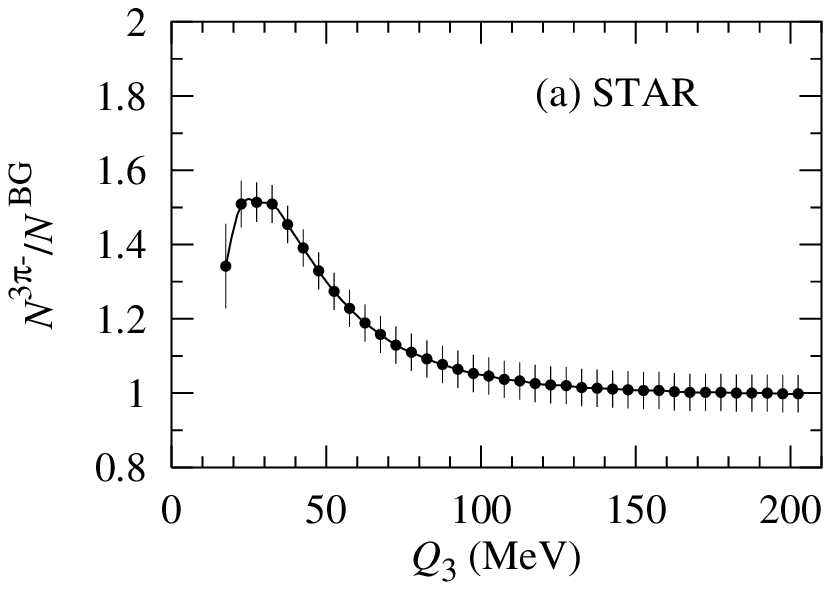}}
\resizebox{0.43\textwidth}{!}{\includegraphics{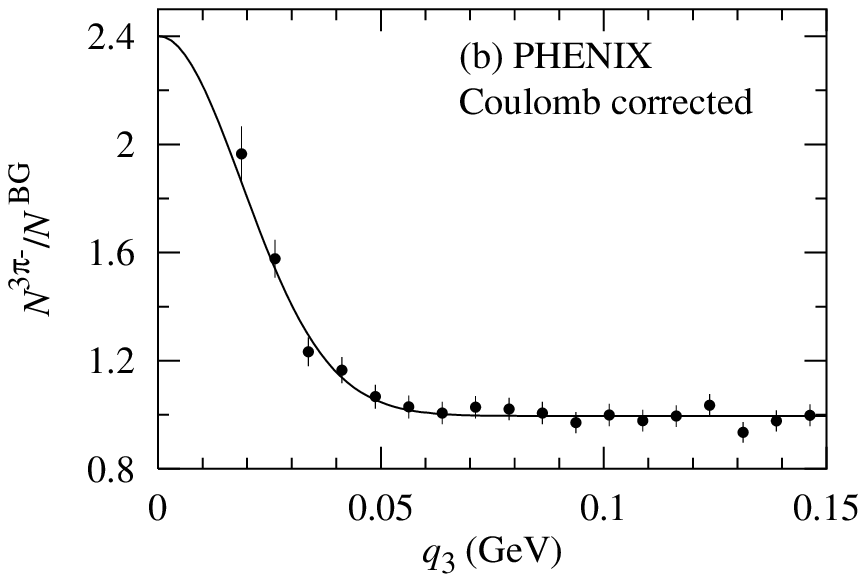}}
\vspace{-5mm}
\caption{Analyses of corrected data by STAR Coll. and PHENIX Coll. Eqs. (\protect \ref{eq04}\protect) and (\protect \ref{eq05}\protect) are used.}
\vspace{-4mm}
\efn{fg05}

{\bf 3-2)} Corrected data with $q_3>0.02$ GeV by PHENIX Coll are analyzed by conventional formula / LO approach. Our results are shown in Figs. \ref{fg05} and \ref{fg07}, and Tables \ref{tb02} and \ref{tb05}.

{\bf 3-3)} Raw data with $q_3 >$ 0.02 GeV are analyzed by the formulae of Coulomb wave function (Eqs. (\ref{eq11}) and (\ref{eq18})). Our results are also shown in Fig. \ref{fg06} and lower-part of Table \ref{tb05}.
\bfn
\vspace{-3mm}
\resizebox{0.40\textwidth}{!}{\includegraphics{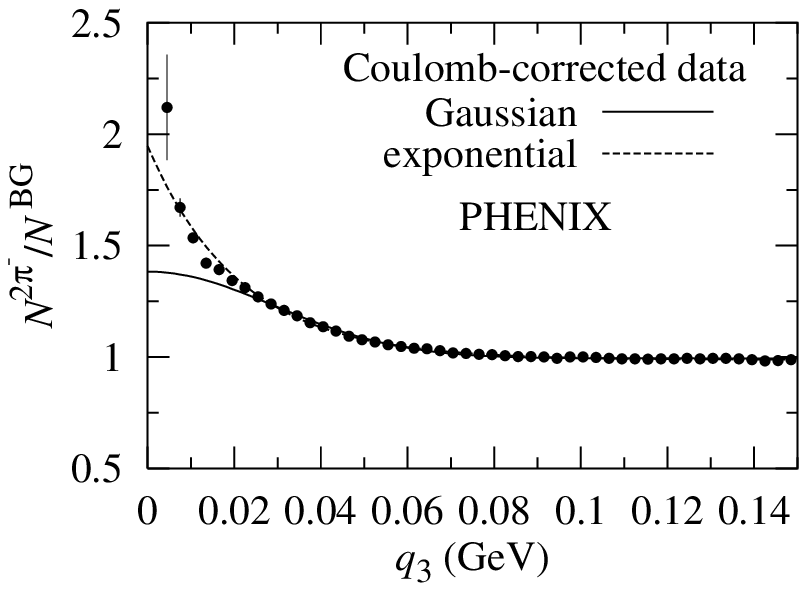}}
\resizebox{0.40\textwidth}{!}{\includegraphics{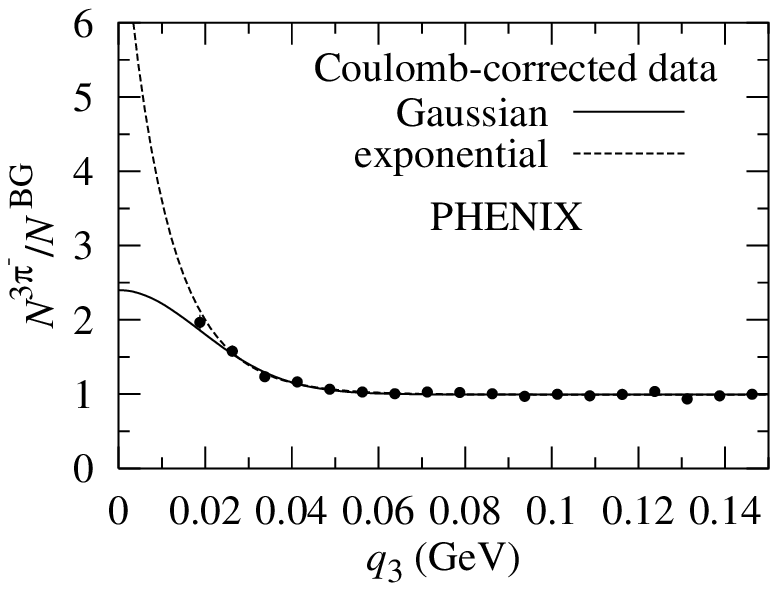}}
\vspace{-5mm}
\caption{Corrected data are analyzed by Eq. (\protect\ref{eq06}\protect) and (\protect\ref{eq07}\protect)}
\vspace{-6mm}
\efn{fg07}
\bfn
\vspace{-3mm}
\resizebox{0.43\textwidth}{!}{\includegraphics{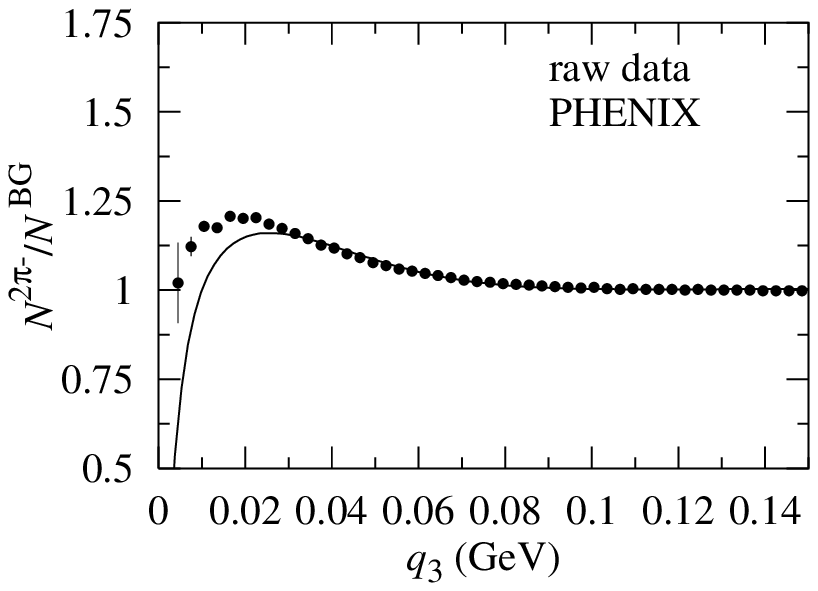}}
\resizebox{0.43\textwidth}{!}{\includegraphics{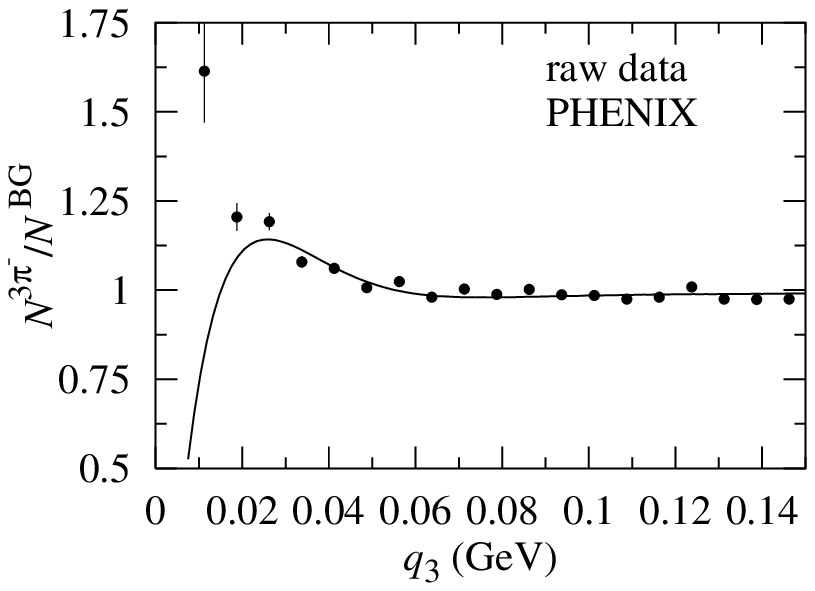}}
\vspace{-5mm}
\caption{Analyses of raw data by PHENIX Coll. Eqs. (\protect \ref{eq11}\protect) and (\protect \ref{eq18}\protect) are used.}
\vspace{-4mm}
\efn{fg06}
\btn
\vspace{-3mm}
\caption{Analyses of data by PHENIX Coll. Eqs. (\protect\ref{eq06}\protect), (\protect\ref{eq07}\protect), (\protect\ref{eq11}\protect) and (\protect\ref{eq18}\protect) are used.}
\begin{tabular}{cc|cccc}
\hline
& $E_{2B}$ & $R$ [fm] & $p$ & $c$ & $\chi^2/N_{dof}$\\
\hline
\lw{$2\pi$} & Gaussian & 6.58$\pm$0.05 & 0.23$\pm$0.00 & 0.98$\pm$0.00 & 156/40 \\
       & Exponential & 9.54$\pm$0.16 & 0.99$\pm$0.01 & 0.99$\pm$0.00 & 56/40 \\
\hline
\lw{$3\pi$} & Gaussian & 9.76$\pm$1.11 & 0.24$\pm$0.08 & 0.99$\pm$0.00 & 7.2/14 \\
       & Exponential & 14.36$\pm$2.10 & 1.00$\pm$0.07 & 0.99$\pm$0.02 & 6.3/14\\\hline
\hline
\multicolumn{2}{c|}{raw data} & $R$ (fm) & $\lambda$ & $c$ & $\chi^2/N_{dof}$\\
\hline
$2\pi$  & Eq. (\protect\ref{eq11}\protect) & 3.77$\pm$0.03 & 0.253$\pm$0.004 & 1.00$\pm$0.00 & 129/40 \\
\hline
$3\pi$  & Eq. (\protect\ref{eq18}\protect) & 5.77$\pm$0.32 & 0.19$\pm$0.02 & 1.00$\pm$0.00 & 84/14 \\
\hline
\end{tabular}
\vspace{-1mm}
\etn{tb05}

{\bf 3-4)} Using two formulae in the core-halo approach (with Gaussian source function, the fraction of core part $f_c$ and the degree of coherent $p_c$ in \cite{Biyajima:2003ey,Biyajima:2005fk,Biyajima:2005ts}) we obtain Fig. \ref{fg09}. In raw data, there is no over-lapping region. On the other hand, in corrected data, we observe very narrow over-lapping region. The reason of the wide region of 3$\pi$ BEC is due to large error bars of corrected data \cite{Csanad:2005nr}.
\bfn
\vspace{-3mm}
\resizebox{0.36\textwidth}{!}{\includegraphics{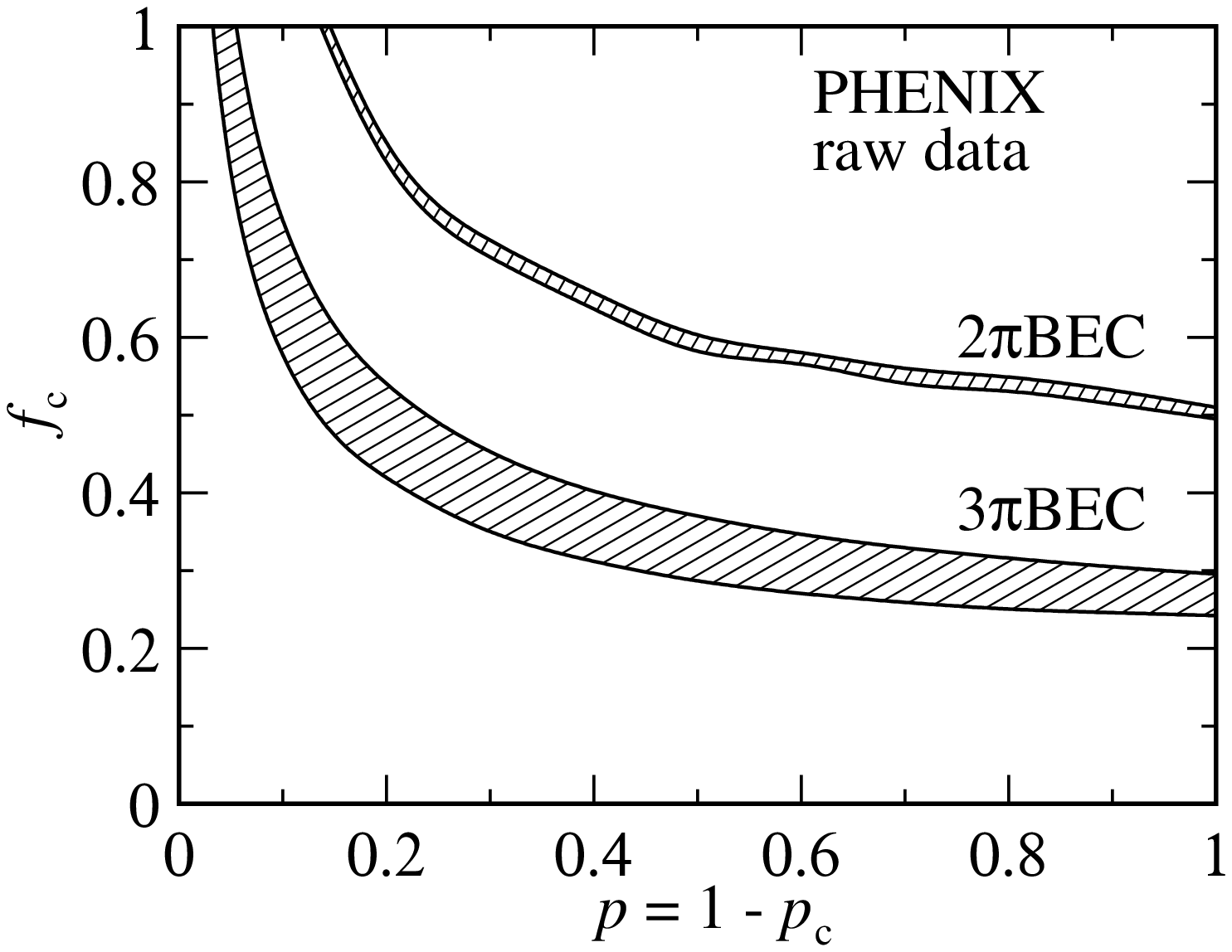}}
\resizebox{0.36\textwidth}{!}{\includegraphics{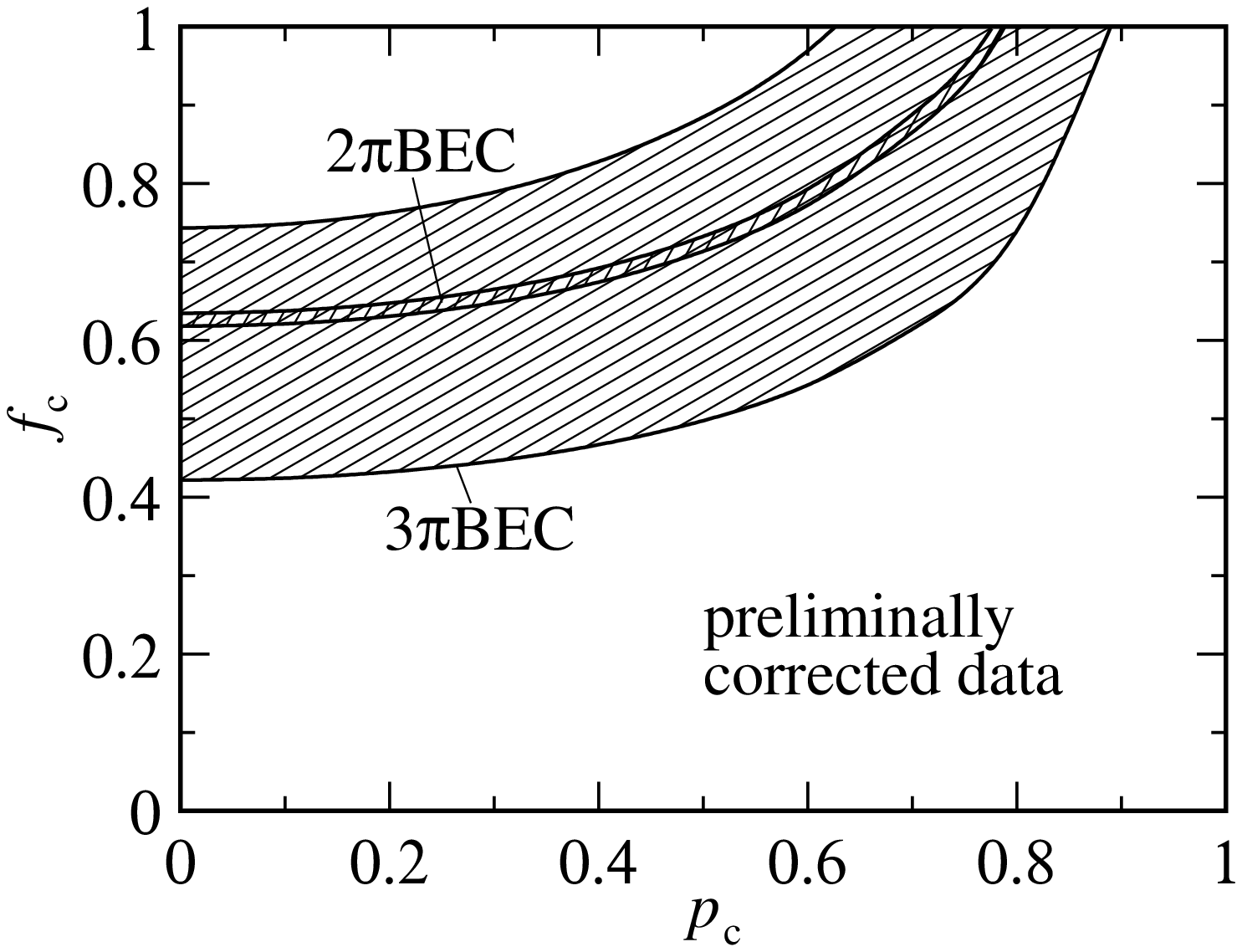}}
\vspace{-3mm}
\caption{Analyses of data by core-halo model.}
\vspace{-6mm}
\efn{fg09}

\vspace{-2mm}
\section{Summary}
\vspace{-2mm}
{\bf 4-1)} From raw data as well as Coulomb corrected data in Au+Au at 130 GeV by STAR Coll., we get the following interaction ranges $R_{2\pi} =$ 8.7 fm and $R_{3\pi} =$ 8.3 fm, and can estimate
\be
V = R_{3\pi}^3\sim 500\ {\rm fm}^3. 
\ee{eq26}
This value is compared with that of ALICE Coll~\cite{Aamodt:2011mr}, $V = R_{\rm long} R_{\rm out} R_{\rm side} \sim 300\ {\rm fm}^3$ at $dN_{ch}/d\eta =$ 1500 and $k_T \sim$ 0.3 GeV in Pb+Pb at 2.76 TeV.

{\bf 4-2)} On the contrary, from corrected data at 200 GeV by PHENIX Coll., we obtain the ranges, by utilizing Eqs. (\ref{eq04}) and (\ref{eq05}), $R_{2\pi} =$ 4.8 fm ($\lambda =$ 0.39) and $R_{3\pi} =$ 6.9 fm ($\lambda =$ 0.34). From raw data, we have smaller interaction ranges $R_{2\pi} = 3.8$ fm and $R_{3\pi} = 5.8$ fm. The interaction ranges of data at 200 GeV by PHENIX Coll. are smaller than those of STAR Coll. At present, it is difficult to draw concrete physical picture for Au+Au collision at 200 GeV. Then we are waiting for final empirical analyses by PHENIX Coll. 

{\bf 4-3)} Moreover, we also eager for empirical analyses of $(2\pi^+)\pi^-$ and $(2\pi^-)\pi^+$ combinations at RHIC and LHC energies. See \cite{Biyajima:2003wg,Abreu:1995sq}.

\end{document}